\documentclass[11pt]{article}
\usepackage{hyperref}
\usepackage{cite}
\usepackage{footnote}
\makesavenoteenv{minipage}
\renewcommand{\title}[1]{%
    \bigskip%
    \begin{center}%
    \Large\bf #1%
    \end{center}%
    \vskip .2in}

\renewcommand{\author}[1]{%
    {\begin{center}
    #1
    \end{center}}}
\newcommand{\address}[1]{\vspace{-1.7em}\vspace{0pt}
    {\begin{center}
    \it #1 
    \end{center}}}

\begin{document}

\title{   BRST symmetry and $W$-algebra in   higher derivative models }

\author
{
Rabin Banerjee $\,^{\rm a,b}$,
Biswajit Paul
$\,^{\rm a,c}$,
Sudhaker Upadhyay $\,^{\rm a, d}$}


\address{ $^{\rm a}${S. N. Bose National Centre 
for Basic Sciences, JD Block, Sector III, Salt Lake City, Kolkata -700 098, India }}
\address{$^{\rm b}$\tt rabin@bose.res.in}
\address{$^{\rm c}$\tt bisu\_1729@bose.res.in}
\address{$^{\rm d}$\tt sudhakerupadhyay@gmail.com }
\begin{abstract}
In this paper we discuss the  (anti-)BRST symmetries and $W$-algebra  of higher derivative theories
of  relativistic particles 
  satisfying general gauge
 conditions.
Using this formalism, the connection  between the (anti-)BRST symmetries and $W$-algebra for the   massless 
 particle with rigidity is established. Incidentally, the full $W$-algebra emerges only when the anti-BRST transformations are considered in tandem with the BRST ones. 
 Further, the BRST symmetry is made finite and
 coordinate-dependent. We show that such finite coordinate-dependent BRST  symmetry
changes the BRST invariant gauge-fixing fermion within the functional integration.
This is exploited to connect two different arbitrary gauge conditions.

\end{abstract}
\section{Introduction}
 It is usual to consider theories where  the Lagrangian has only single time derivative of the fields. But in some cases  we need to consider terms where higher time derivative of the fields appear. Such theories are known as  higher derivative (HD) theories. The concept of introduction of HD field is not new and has been considered by many authors and applied in diverse fields like electrodynamics \cite{podolsky1, podolsky2}, supersymmetry   \cite{Iliopoulos, Gama}, noncommutativive theory      \cite{clz, plyuschay6},  cosmology     \cite{neupane, nojiri4}, extended Maxwell-Chern-Simon theory \cite{reyes, MP}, theory of anyons\cite{plyuschay3, plyuschay4, plyuschay5}, relativistic particle with torsion \cite{plyuschay7}, membrane model of the electron \cite{cordero, paul} etc. In gravity theories HD terms were added to ensure renormalizability \cite{stelle}. There are various models of gravity where HD corrections are added to the Einstein-Hilbert action\cite{accioly, soti, gullu, ohta}. HD terms also frequently appear in the context of string theory \cite{polya, elie}. The importance of HD terms, therefore, cannot be overemphasised. A useful and interesting HD model to be considered is  the relativistic particle model with curvature. In this case the curvature term, which is higher derivative in nature, is added to the action of the usual massive relativistic particle. This model was introduced long ago by Pisarski  \cite{pisarski} and still continues to be under active consideration \cite{nesterenko, plyuschay1, plyuschay2, plyuschayA, ramos, ramos2, BMP}. Interestingly, the model has only one gauge symmetry identified as diffeomorphism symmetry although there are two independent primary first class constraint present in the theory \cite{BMP},  which is unusual. The presence of the extra primary first class constraint was successfully explained as an effect of the higher derivative nature.    The massless version, known as massless 
 particle model with rigidity, is shown to describe bosons and fermions  \cite{plyuschay2}. This model has two gauge symmetries viz. diffeomorphism and W-symmetry alongwith two primary first class constraints \cite{ramos, ramos2, BMP}. One can also add a torsion term to the relativistic particle model and the theory emerges with very interesting results. When quantised, the relativistic particle model with torsion leads to Majorana equations\cite{plyuschay7}. The massive sector contains infinite number of states when quantised in the Minkowski space and finite number of states in euclidean space \cite{plyuschay3}. Higher derivative models also shows Majorana equations \cite{plyuschay5} with the 2+1 dimensional analogous models lead to anyons. Another interesting line where the higher derivative models appear corresponds to finite-gap (algebra-geometric) systems\cite{novikov}. There, the higher derivative models are given by the so called Novikov equation, with a space variable playing a role of the
evolution parameter.

On the other hand, BRST symmetry is a very powerful tool to quantize a theory with gauge invariance which also 
helps 
in the proof of the renormalizability and  unitarity of  gauge  theories \cite{brst, tyu,ht,wei}. 
This transformation, which is characterized by an infinitesimal,
global   and
anticommuting parameter leaves the effective action as well as path integral of the effective theory invariant.
In  gauge field theories the usual BRST symmetry has been generalized 
 to make it finite and field-dependent \cite{sdj}. 
This finite field dependent BRST (FFBRST) transformations have found many applications in various contexts
\cite{sb1,jm, ssb, susk, sb, rb, smm, fs}.

  The implementation of BRST symmetries for HD theories is quite nontrivial and poses problems. In this context, therefore, a natural question arises regarding  the application of BRST formalism to relativistic particle models. Indeed it is not surprising that in spite of a considerable volume of research on relativistic particle models, this aspect remains unstudied. A basic motivation of this paper is to bridge this gap.
    
In this paper, we consider a relativistic particle model with curvature as a HD theory  possessing a gauge 
symmetry. The constraint analysis of this model  and its  massless analogue is discussed. Further,
the gauge symmetry transformations in the case of massive relativistic particle model with curvature are identified with diffeomorphism  invariance.  However, for the  massless 
 particle model with rigidity
it corresponds to $W$-symmetry in addition to diffeomorphism invariace. 
We  construct the BRST symmetry and anti-BRST symmetry for these particle models. The difficulties of applying BRST transformations to HD theories are bypassed by working in the first order formalism developed in \cite{plyuschay1, plyuschay2, BMP} instead of the conventional Ostrogradski approach \cite{ostro}.
It is shown that the (anti-)BRST symmetry transformations for all variables reproduce  the 
diffeomorphism symmetry of the massive relativistic particle model including curvature. Furthermore, we also show that the  massless 
 particle model with rigidity  yields both the diffeomorphism and   $W$-invariances. We explicitly demonstrate  the $W_3$-algebra.
For BRST transformations this algebra is shown for all variables,
excluding the anti-ghost. Exactly the same features are revealed, but now excluding the ghost variable instead of the anti-ghost, for anti-BRST transformations. To get the complete picture, therefore, both BRST and anti-BRST transformations have to be considered.
Further, we implement the concept of FFBRST transformation \cite{sdj} in the quantum mechanical relativistic particle model.
The quantum mechanical version of FFBRST transformation \cite{sdj}
is called as finite coordinate-dependent BRST (FCBRST) transformation.  
We see that FCBRST transformation for the relativistic particle model
 is a symmetry of the action only, but not of the 
generating functional. Analogous to FFBRST the FCBRST transformation changes the Jacobian of path integral
measure non-trivially. For an appropriate choice of finite coordinate dependent parameter
FCBRST connects two different gauge-fixed action within functional integration. 

The plan of the paper is as follows. In sections 2 and 3 we introduce the various relativistic particle models and  discuss their gauge  symmetries. The nilpotent BRST and anti-BRST transformation
with emergence of $W_3$-algebra is demonstrated in section 4. In section 5, we construct 
 the FCBRST transformation. Two arbitrary gauges are connected with the help of this
FCBRST  within a path integral formalism in section 6. We draw concluding remarks in the last section.

\section{Massive relativistic particle model with curvature}

The massive relativistic point particle theory with curvature   has the action\footnote{contractions are abbreviated as $A^\mu B_\mu = AB$, $A^\mu A_\mu=A^2$. We consider the model in 3 + 1 dimensions. So $\mu$ assumes the values 0, 1, 2, 3 \cite{plyuschay1, plyuschayA, BMP}.} 
\begin{equation}
S=-m \int{\sqrt{\dot{x}^{2}}}d\tau+\alpha\int{\frac{\left( \left( \dot{x}\ddot{x}\right)^{2}-\dot{x}^{2}\ddot{x}^{2} \right)^\frac{1}{2}}{\dot{x}^{2}}}d\tau.\label{maction}
\end{equation}
The model is meaningful for  $\alpha < 0 $ and $\dot{x}^2 > 0$ . The energy spectrum for this model turns out to be the energy spectrum of the  Majorana equation \cite{plyuschay1, plyuschayA}. 
 By direct substitution we may verify that  (\ref{maction}) is invariant under reparametrisation,
\begin{equation}
\tau \longrightarrow \tau + \Lambda(\tau),
\end{equation}
where $\Lambda$ is an infinitesimal reparametrisation parameter. Under this reparametrisation $x^{\mu}$ transforms as,
\begin{equation}
\delta x^{\mu} = x^{\mu}(\tau - \Lambda) - x^{\mu}(\tau) =-\Lambda \dot{x}^{\mu}.
\label{delx}
\end{equation} 
Since this is a higher derivative model, the usual Hamiltonian formalism does not apply. There is, however,  the well established Ostrogradski method where the momenta are defined in some non-trivial way \cite{ostro}. Other than this, a first order formalism exists in the literature where the time  derivatives of the coordinates are considered as independent variables  to convert the theory into a first order one. There are different variants \cite{plyuschay1, plyuschay2, BMP} of this formalism and we adopt the one that was developed by two of us in a collaborative work \cite{BMP}. To convert the theory into a first order one we introduce the new coordinates
\begin{equation}
q^{\mu}_1 = x^{\mu}\ \;\ \ q^{\mu}_2 = \dot{x}^{\mu}.
\label{mnewcoordinate}
\end{equation}
The Lagrangian in these coordinates has a first order form given by 
\begin{eqnarray}
L &=& -m\sqrt{q_{2}^{2}} + \alpha \frac{\left({\left({q_{2}\dot{q}_{2}} \right)^{2} - q_{2}^{2} \dot{q}_{2}^{2} } \right)^{\frac{1}{2}} }{q_{2}^{2}} + q_{0}^{\mu}(\dot{q}_{1\mu} - q_{2\mu}),
\label{mindependentfields}
\end{eqnarray}
where $q_{0}^{\mu}$ are the Lagrange multipliers that enforce the constraints
\begin{equation}
\dot{q}_{1\mu} - q_{2\mu} = 0.
\label{27}
\end{equation}
Let $p_{0\mu}$, $p_{1\mu}$ and $p_{2\mu}$ be the canonical momenta conjugate to $q_{0\mu}$, $q_{1\mu}$ and $q_{2\mu}$ respectively,
\begin{eqnarray}
 p_{0\mu} &=& \frac{\partial{L}}{\partial{\dot{q}_{0}^{ \ \mu}}} =0, 
 \nonumber\\ 
 p_{1\mu} &=& q_{0\mu},
 \nonumber\\
 p_{2\mu} &=& \frac{\alpha ((q_{2} \dot{q}_{2})q_{2\mu}  - q_{2}^{2} \dot{q}_{2\mu} ) } {q_{2}^{2} \sqrt{(q_{2} \dot{q}_{2})^{2} - q_{2}^{2} \dot{q}_{2}^{2}}}.
 \label{momenta}
  \end{eqnarray}
 The primary  constraints thus obtained are listed below \cite{plyuschayA, BMP}, 
\begin{eqnarray}
\Phi_{0\mu} &=&  p_{0\mu} \approx 0, \nonumber\\
\Phi_{1\mu} &=& p_{1\mu} - q_{0\mu} \approx 0,
\nonumber\\
\Phi_{1} &=& p_{2}q_{2} \approx 0, \nonumber\\
\Phi_{2} &=& p_{2}^{2}q_{2}^{2} + \alpha^{2} \approx 0.
\label{mPFC2}
\end{eqnarray}
Conservation of the constraints (\ref{mPFC2}) yield  the following secondary constraints, 
\begin{eqnarray}
\omega_{1} &=&  q_{0}q_{2} +  m\sqrt{q_{2}^{2}} \approx 0, \nonumber\\
\omega_{2} &=& q_{0}p_{2} \approx 0.
\label{mssc}
\end{eqnarray}
We define $\Phi_{2}^{\prime}$  as a combination of constraints,
\begin{equation}
\Phi_{2}^{\prime}=\left(q_{0}^{2}-m^{2} \right)\Phi_{2} - 2p_{2}^{2}\left( q_{0}q_{2}\right)\omega_{1}.
\label{44}  
\end{equation}
 $\Phi_{1}$ and $\Phi_{2}^{\prime}$ form the first class constraint set which are primary in nature. The second class constraints $\Phi_{0\mu}$, $\Phi_{1\mu}$, $\omega_{1}$ and $\omega_{2}$ are eliminated by   replacing all the Poisson brackets by Dirac brackets. The nonzero Dirac brackets between the phase space variables are listed below\footnote{ Dirac brackets are denoted by $\{ , \}_D$. We consider  $p^{2}_{1} - m^{2} \neq 0$, else it is a singular case. Explicit constraint structure and Dirac brackets of the singular case  can be found in \cite{BMP}. }
 \begin{eqnarray}
\nonumber
\{ q_{1\mu}, q_{1\nu}\}_{D}
&=& \frac{1}{p_{1}^{2}-m^{2}}\left({p_{2\mu}q_{2\nu}-q_{2\mu}p_{2\nu}} \right),     
\nonumber\\
\{ {q_{1\mu}, q_{2\nu}}\}_{D}
&=&-\frac{q_{2\mu}p_{1\nu}}{p_{1}^{2}-m^{2}},      
\nonumber\\
\{{q_{1\mu}, p_{2\nu}} \}_{D}
&=& \frac{1}{p_{1}^{2}-m^{2}}\left( {\frac{m}{\sqrt{q_{2}^{2}}}p_{2\mu}q_{2\nu} + p_{2\mu}p_{1\nu}}\right),
\nonumber \\
\{q_{1\mu}, p_{1\nu}\}_{D} &=& \eta_{\mu\nu},
\nonumber \\
\{{q_{2\mu}, p_{2\nu}} \}_{D}&=& \eta_{\mu\nu}-\frac{1}{p_{1}^{2}-m^{2}}\left({p_{1\mu}p_{1\nu} + \frac{m}{\sqrt{q_{2}^{2}}}p_{1\mu}q_{2\nu}} \right).
\label{50} 
\end{eqnarray}
Now the generator $G$ of the gauge symmetry is given by a combination of the first class constraints,
\begin{equation}
G = \epsilon^{1} \Phi_{1} + \epsilon ^{2}(p_{1}^{2} - m^2)\Phi_{2}.
\end{equation}
However, we have shown in \cite{BMP} that there is only one independent gauge parameter which is a consequence of the higher derivative theory. The generator is then given by, 
\begin{equation}
G= \frac{q_{2\mu}}{q_{2}^{2}}\frac{d}{d\tau}\left(2mp_{2}^{2}\sqrt{q_{2}^{2}}\epsilon^{2}q_{2}^{\mu} \right)\Phi_{1} + \epsilon^{2}(p_{1}^{2} - m^2)\Phi_{2}.
\label{generator}
\end{equation}

 It may be noted that in the expression of the gauge generator there appears time derivative of the fields as well gauge parameter $\epsilon^2$. The apparent problem of taking gauge variation of the derivative of the phase space variables does not arise  since they are multiplied by constraints which are set to zero after computing the brackets.   Now the gauge transformation of the variables are given by 
\begin{eqnarray}
\delta{q_{1}^{\mu}} =\{q_{1}^{\mu}, G\}_{D}=
 \left( 2\epsilon^{2}p_{2}^{2}m\sqrt{q_{2}^{2}}\right)  q_{2}^{\mu},
\end{eqnarray}
and
\begin{eqnarray}
\delta{q_{2}^{\mu}} = \{q_{2}^{\mu}, G\}_{D} =  \left[ \frac{q_{2\nu}}{q_{2}^{2}}\frac{d}{d\tau}\left(2mp_{2}^{2}\sqrt{q_{2}^{2}}\epsilon^{2}q_{2}^{\nu} \right)\right] q_{2}^{\mu}+ \left[ 2\epsilon^{2}q_{2}^{2}(p_{1}^{2} - m^{2})\right] p_{2}^{\mu}.
\end{eqnarray}
 In terms of   the reparametrization parameter,  $\Lambda = -2\epsilon^{2}p_{2}^{2}m\sqrt{q_{2}^{2}}$,  
  the transformation for $q_{1}^{\mu}$ may be expressed as,
\begin{eqnarray}
\delta{q_{1}^{\mu}} = -\Lambda q_2^\mu.
 \label{cmp1}
\end{eqnarray}
Using the identification (\ref{mnewcoordinate}) the relation (\ref{cmp1}) reproduces the reparametrisation symmetry (\ref{delx}). Thus the gauge symmetry gets identified with the reparametriasation symmetry.

\section{The model of massless particle with rigidity}

	 The massless version of the model (\ref{maction})  is known as the model of massless particle with rigidity. The massless version is not obtained  simply by putting $m=0$ in (\ref{maction}) due to reasons of internal consistency \cite{plyuschay2}. This requires a modification in the curvature term and the model is given by, 
\begin{equation}
S=\alpha\int{\frac{ \left(\dot{x}^{2}\ddot{x}^{2} -  \left(\dot{x}\ddot{x}\right)^{2} \right)^\frac{1}{2}}{\dot{x}^{2}}}d\tau.\label{action}
\end{equation}
 The action of this model thus is proportional to the curvature and also its classical equation of motion is compatible only for super-relativistic motion of the particle. The importance of the model lies in the fact that  it corresponds to massless modes of either integer or half-integer helicity states. The model finds its relevance when $\dot{x}^2 < 0$ \cite{plyuschay2}. Once again we adopt  the first order formalism developed in \cite{BMP}. We introduce the new coordinates
\begin{equation}
q^{\mu}_1 = x^{\mu},\ \  \ \ q^{\mu}_2 = \dot{x}^{\mu}.
\label{newcoordinate}
\end{equation}
The Lagrangian in these coordinates has a first order form given by
\begin{eqnarray}
L &=& \alpha \frac{\left({q_{2}^{2} \dot{q}_{2}^{2} -\left({ q_{2}\dot{q}_{2}} \right)^{2}  } \right)^{\frac{1}{2}} }{q_{2}^{2}} + q_{0}^{\mu}(\dot{q}_{1\mu} - q_{2\mu}),
\label{independentfields}
\end{eqnarray}
where $q_{0}^{\mu}$ are the Lagrange multipliers that enforce the same constraints mentioned in equation (\ref{27}).
 
Let $p_{0\mu}$, $p_{1\mu}$ and $p_{2\mu}$ be the canonical momenta conjugate to $q_{0\mu}$, $q_{1\mu}$ and $q_{2\mu}$ respectively, having same expressions as that of (\ref{momenta}). Consequently, we obtain the following  primary constraints 
\begin{eqnarray}
\nonumber
\Phi_{0\mu} &=& p_{0\mu} \approx 0,
\nonumber \\
\Phi_{1\mu} &=& p_{1\mu} - q_{0\mu} \approx 0,
\nonumber\\
\nonumber
\Phi_{1} &=& p_{2}q_{2} \approx 0,
\nonumber \\
\Phi_{2} &=& p_{2}^{2}q_{2}^{2} - \alpha^{2} \approx 0.
\label{PFC2new}
\end{eqnarray}
The secondary set of constraints obtained by time conserving the primary ones are,
\begin{eqnarray}
\nonumber
\omega_{1} &=&  q_{0}q_{2}  \approx 0,
\nonumber \\
\omega_{2} &=& q_{0}p_{2} \approx 0.
\end{eqnarray}
Finally, by conserving $\omega_{2}$ the tertiary constraint is obtained as   
\begin{eqnarray}  
\omega_{3} &=& q_{0}^{2} \approx 0.
\label{36new}
\end{eqnarray} 
This completes the chain of constraints. In the above constraint structure the first class set is \{ $\Phi_{1}$,$\Phi_{2}$, $\omega_{1}$, $\omega_{2}$, $\omega_{3}$\} and all others are second class in nature. Once again we remove all second class constraints as described in the previous section. Fortunately the Dirac bracket comes out to be same as the Poisson brackets between the phase space variables.  
After removing the nondynamical variables $q_{0\mu}$ and $p_{0\mu}$ by solving the second class constraints $\Phi_{0\mu}$ and $\Phi_{1\mu}$,  the final set of first class constraints  are

\begin{eqnarray}
\Omega_1 &=& \Phi_{1} = p_{2}q_{2} \approx 0,\nonumber\\
\Omega_2  &=& \Phi_{1} =  p_{2}^2q_{2}^2 - \alpha^2 \approx 0,\nonumber\\
\Omega_3  &=& \omega_{1} = p_1q_2\approx 0,\nonumber\\
\Omega_4  &=& \omega_{2} = p_1p_2\approx 0,\nonumber\\
\Omega_5  &=& \omega_{3} = p_1^2 \approx 0.
\end{eqnarray}
Note that the original condition $\dot{x}^2 < 0$ translates into $q_{2}^{2} < 0$. This ensures the reality of $\alpha$, as may be easily seen from the constraint $\Omega_2 \approx 0$. The reality of $\alpha$ is connected to the helicity states of the particles as discussed \cite{plyuschay2}.\\
 As done previously, the gauge generator is written as a combination of all the first class constraints,
\begin{equation}
G = \sum_{a=1}^{5} \epsilon^{a}\Omega_{a}.
\label{rigidsumG}
\end{equation} 
 However, due to the presence of secondary first-class constraints, the parameters of gauge transformation($\epsilon^{a}$) are not independent \cite{ht, BRR}. It is found that only two out of the five gauge parameters are independent.For our  convenience we  take $\epsilon^3$ and $\epsilon^5$ as independent.

So, the expression for the gauge generator becomes \cite{BMP},
\begin{eqnarray}
G &=& \left( {\dot{\epsilon}}^3  + \frac{q_2{\dot{q}}_2}{q_2^{\ 2}}\epsilon^3 - \frac{\alpha\sqrt{g}}{q_2^{\ 4}}\dot{\epsilon}^{5}\right) \Omega_{1}+  \frac{1}{2q_2^{\ 2}}
\left({\ddot{\epsilon}}^5 - \frac{q_2{\dot{q}}_2}{q_2^{\ 2}}{\dot{\epsilon^5}} + \frac{\alpha\sqrt{g}}{p_2^{\ 2}q_2^{\ 2}}\epsilon^3\right)\Omega_{2}\nonumber\\
&& 
+\epsilon^{3}\Omega_{3}+\dot{\epsilon}^{5}\Omega_{4}+\epsilon^{5}\Omega_{5},
\end{eqnarray}
which contains only two independent parameters (here $g=q_{2}^{2} \dot{q}_{2}^{2} -\left({ q_{2}\dot{q}_{2}} \right)^{2}$). 
	
	 We now calculate the gauge variations of the dynamical variables, defined as $\delta{q} = \left\lbrace {q, G}\right\rbrace_D $. These are given by,
\begin{eqnarray}
\nonumber
\delta{q_{1}^{\mu}} &=& \epsilon^{3}q_{2}^{\mu}+ \dot{\epsilon}^{5}p_{2}^{\mu} + 2\epsilon^{5}p_{1}^{\mu},\\
\nonumber
\delta{q_{2}^{\mu}} &=& \left( {{\dot{\epsilon}}^3  + \frac{q_2{\dot{q}}_2}{q_2^{\ 2}}\epsilon^3 - \frac{\alpha\sqrt{g}}{q_2^{\ 4}}\dot{\epsilon}^5 } \right)  q_{2}^{\mu} + \left({\ddot{\epsilon}}^5 - \frac{q_2{\dot{q}}_2}{q_2^{\ 2}} {\dot{\epsilon^5}} + \frac{\alpha\sqrt{g}} {p_2^{\ 2}q_2^{\ 2}}\epsilon^3\right)p_{2}^{\mu} + \dot{\epsilon}^{5}p_{1}^{\mu},\\
\nonumber
\delta{p_{1}^{ \mu}} &=& 0,\\
\nonumber
 \delta{p_{2}^{\mu}} &=& -\left( {{\dot{\epsilon}}^3  + \frac{q_2{\dot{q}}_2}{q_2^{\ 2}}\epsilon^3 - \frac{\alpha\sqrt{g}}{q_2^{\ 4}}\dot{\epsilon}^5} \right)  p_{2}^{\mu} \nonumber\\
 &&
 -\frac{p_{2}^{2}} {q_{2}^{2}} \left({\ddot{\epsilon}}^5 - \frac{q_2{\dot{q}}_2}{q_2^{\ 2}} {\dot{\epsilon^5}} + \frac{\alpha\sqrt{g}} {p_2^{\ 2}q_2^{\ 2}}\epsilon^3\right)q_{2}^{\mu}  -\epsilon^{3}p_{1}^{\mu}.
\label{dwinvariance}
\end{eqnarray}

  The above  transformations  can be identified as diffeomorphism ($D$) and  $W$-symmetry by putting $\epsilon^{5} = 0$ and $\epsilon^{3} = 0$ respectively, as done in \cite{BMP}. Detailed calculations on all the phase-space variables show that\cite{ramos, ramos2, BMP}
 \begin{eqnarray}
 \nonumber
\left[ { \delta^{D}_{\epsilon^{3}_{1}} , \delta^{D}_{\epsilon^{3}_{2}} }\right]   &=& \delta^{D}_{\epsilon^{3}}; \ \ \ \mbox{with}  \ \ \ \ \epsilon^{3}= \dot{\epsilon}_{1}^{3} \epsilon_{2}^{3} - \epsilon_{1}^{3} \dot{\epsilon}_{2}^{3}  \\
 \nonumber
\left[ { \delta^{D}_{\epsilon^{3}} , \delta^{W}_{\epsilon^{5}}}\right]  &=& \delta^{W}_{\epsilon^{\prime 5}}; \ \ \ \mbox{with}  \ \ \ \ \epsilon^{\prime5}= -\epsilon^{3}\dot{\epsilon}^{5}\\ 
 \nonumber
\left[ { \delta^{W}_{\epsilon^{5}_{1}} , \delta^{W}_{\epsilon^{5}_{2}} }\right]   &=& \nonumber \delta^{W}_{\epsilon^{5}}; \ \ \ \mbox{with}  \ \ \ \ \epsilon^{5}= \frac{p_{2}^{2}}{q_{2}^{2}}\left( \dot{\epsilon}_{2}^{5} \epsilon_{1}^{5} - \epsilon_{2}^{5} \dot{\epsilon}_{1}^{5}  \right). 
\end{eqnarray}  
This reproduces the usual  $W_3$-algebra.

\section{(Anti-)BRST symmetries and $W_3$-algebra }
 In this section we construct the    nilpotent BRST  and anti-BRST  symmetries for the theory. For this purpose
we  need to fix a gauge before the quantization of the theory as the theory is gauge invariant and therefore has some redundant degrees of freedom. The
general gauge condition in this case is chosen as:
\begin{equation}
 F_1 [f(q)] =0, \label{gauge1}
\end{equation}
where $f(q)$ is a general function of all the generic variables $q$. Some 
explicit
examples of gauge  conditions corresponding to
relativistic particle models 
 are \cite{plyuschay2}.
 \begin{eqnarray}
q_1^0 -\tau =0,\ \  q_2^0 -1 =0,\ \  p_2^0 =0,\ \ q_2^2 =0.
 \end{eqnarray}
The general gauge condition (\ref{gauge1}) can be incorporated at a quantum level by adding the 
appropriate gauge-fixing term to  classical action. 
 
The linearised gauge-fixing term  using
Nakanishi-Lautrup auxiliary variable $  B ( q) $ is given by
\begin{equation}
 S_{gf} = \int d\tau    \left [ \frac{1}{2} B^2 +BF_1 [f(q) ] \right].
\end{equation}
To complete the effective theory we need a further Faddeev-Popov ghost term in the action. The ghost term in this  case is constructed as    
\begin{eqnarray}
  S_{gh} &=&  \int d\tau    \left [  \bar c s F_1 [f(q) ] \right],\nonumber\\
 &=& - \int d\tau    \left [    c \bar s F_1 [f(q) ] \right], 
\end{eqnarray}
where 
  $c$ and $ \bar{c}  $ are ghost and  anti-ghost variables.  
 Now the effective action can be written as
 \begin{equation}
 S_{eff} =S+S_{gf} +S_{gh}.\label{seff}
 \end{equation}
 The source free generating functional for this theory is defined as
 \begin{equation}
 Z[0] =\int {\cal D} q\ e^{iS_{eff}},\label{zfun}
 \end{equation}
 where ${\cal D} q$ is the path integral measure. 
 The nilpotent BRST symmetry of the  effective action in the case of relativistic 
 particle model with curvature 
is defined  by  replacing the infinitesimal reparametrisation parameter ($\Lambda$) to ghost 
variable $c$ in the gauge transformation given in equation (\ref{cmp1}) as 
 \begin{eqnarray}
&& s^{D} q_1^\mu =-c  q_2^\mu,\ \ \ s^{D} q_2^\mu =-\dot c   q_2^\mu -c  \dot q_2^\mu,\nonumber\\
&&  s^{D} c =0,\ \ \ s^{D} \bar c  =B, \ \ \ s^{D} B  =0,
\label{brs1}
 \end{eqnarray}
 where $c$, $\bar c $ and $B $ are ghost, anti-ghost and auxiliary variables  respectively 
 for relativistic  particle model with curvature.  This  BRST transformation, corresponding to
 gauge symmetry identified with
 the diffeomorphism invariance, leaves both  the effective action as well as 
 generating functional, invariant. 
 Similarly, we construct the anti-BRST symmetry
 transformation, where the roles of ghost and anti-ghosts are interchanged with some coefficients,
 as 
  \begin{eqnarray}
&&\bar s^{D} q_1^\mu =-\bar c  q_2^\mu,\ \ \ \bar  s^{D} q_2^\mu =-\dot {\bar c}   q_2^\mu -\bar c  \dot q_2^\mu,
\nonumber\\
&& \bar s^{D} 
\bar c =0,\ \ \ \bar s^{D}  c  =-B, \ \ \ \bar s^{D} B  =0.\label{antibrs1}
 \end{eqnarray}
 These transformations are nilpotent and absolutely anticommuting in nature i.e.
 \begin{equation}
 (s^{D})^{2}=0,\ \ ( {\bar s^D})^2=0,\ \ s^{D}\bar s^{D}+\bar s^{D} s^{D} =0.
 \end{equation}
 The above (anti-)BRST transformations are  valid for both the models. On the other hand, the nilpotent BRST and anti-BRST symmetry transformations,  identified with
 $W$-symmetry (with $\epsilon^3 =0$) in (\ref{dwinvariance}), for relativistic massless particle model with rigidity  only, are constructed as  
  \begin{eqnarray}
 s^{W} q_1^\mu &=& \dot \eta p_2^\mu +2 \eta p_1^\mu,\nonumber\\
 s^{W} {q_{2}^{\mu}} &=&  - \frac{\alpha\sqrt{g}}{q_2^{\ 4}}\dot \eta   q_{2}^{\mu} +\left({\ddot \eta}  - \frac{q_2{\dot{q}}_2}{q_2^{\ 2}} {\dot \eta} \right)p_{2}^{\mu} + \dot \eta p_{1}^{\mu},
 \nonumber\\
 s^{W} p_1^\mu &=&0, \ \ s^{W} p_2^\mu =    \frac{\alpha\sqrt{g}}{q_2^{\ 4}}\dot \eta  p_{2}^{\mu} -\frac{p_{2}^{2}} {q_{2}^{2}}\left({\ddot \eta}  - \frac{q_2{\dot{q}}_2}{q_2^{\ 2}} {\dot \eta} \right)q_{2}^{\mu},\nonumber\\
      s^{W} \eta &=&0,\ \ s^{W} \bar \eta =B, \ \ s^{W} B =0,\label{brs2}
 \end{eqnarray} and
  \begin{eqnarray}
\bar s^{W} q_1^\mu &=& \dot {\bar \eta} p_2^\mu +2 {\bar \eta} p_1^\mu,\nonumber\\
 \bar s^{W} {q_{2}^{\mu}} &=&  - \frac{\alpha\sqrt{g}}{q_2^{\ 4}}\dot {\bar \eta}   q_{2}^{\mu} +\left({\ddot{\bar\eta}}  - \frac{q_2{\dot{q}}_2}{q_2^{\ 2}} {\dot {\bar\eta}} \right)p_{2}^{\mu} + \dot {\bar \eta} p_{1}^{\mu},
 \nonumber\\
\bar s^{W} p_1^\mu &=&0, \ \ \bar s^{W} p_2^\mu =    \frac{\alpha\sqrt{g}}{q_2^{\ 4}}\dot{\bar \eta}  p_{2}^{\mu} -\frac{p_{2}^{2}} {q_{2}^{2}}\left({\ddot {\bar \eta}}  - \frac{q_2{\dot{q}}_2}{q_2^{\ 2}} {\dot{\bar \eta}} \right)q_{2}^{\mu},\nonumber\\
  \bar    s^{W} {\bar \eta}&=&0,\ \ \bar s^{W} \eta =-B, \ \ \bar s^{W} B =0,
  \label{antibrs2}
 \end{eqnarray}
  where $\eta, \bar \eta$ and $B $ are ghost, anti-ghost and auxiliary variables,  respectively, 
 for  relativistic massless particle model with rigidity. 
 
 Here we observe interestingly  that the BRST symmetry transformations of all variables 
(excluding the anti-ghost  variable) given in equations (\ref{brs1}) and (\ref{brs2}) also satisfy the $W_3$-algebra as
  \begin{eqnarray}
 \nonumber
\left[ {s_{c_1}^{D} } , s_{c_2}^{D}  \right]   &=& s_{c_3}^{D};\ \ \ \mbox{with}  \ \ \ \ c_3= c_2 \dot c_1 - \dot c_2 c_1\\ 
 \nonumber
\left[ { s_c^{D}, s_\eta^{W}}\right]     &=& s_{\eta'}^{W};\ \ \ \mbox{with}  \ \ \ \ \eta' =\dot \eta c\\ 
 \nonumber
\left[ { s_{\eta_1}^{W}, s_{\eta_2}^{W} }\right]     &=& \nonumber s_{\eta_3}^{W};
\ \ \ \mbox{with}  \ \ \ \ \eta_3= \frac{p_{2}^{2}}{q_{2}^{2}}\left( \eta_2\dot\eta_1 -\eta_1\dot \eta_2 \right),
\end{eqnarray} 
and the anti-BRST symmetry transformations of all variables 
(excluding  ghost  variable) given in equations (\ref{antibrs1}) and (\ref{antibrs2}) also satisfy the $W_3$-algebra as
  \begin{eqnarray}
 \nonumber
\left[ {\bar s_{\bar c_1}^{D} } , \bar s_{\bar c_2}^{D}  \right]    &=& \bar s_{\bar c_3}^{D};\ \ \ \mbox{with}  \ \ \ \ \bar c_3= \bar c_2 \dot {\bar c}_1 - \dot {\bar c}_2 \bar c_1 \\
 \nonumber
\left[ {\bar s_{\bar c}^{D}, \bar s_{\bar \eta}^{W}}\right]     &=&\bar  s_{\bar \eta'}^{W};\ \ \ \mbox{with}  \ \ \ \ \bar\eta' =\dot {\bar\eta}\bar c\\ 
 \nonumber
\left[ {\bar s_{\bar\eta_1}^{W}; \bar s_{\bar\eta_2}^{W} }\right]     &=& \nonumber \bar s_{\bar\eta_3}^{W};\ \ \ \mbox{with}  \ \ \ \ \bar\eta_3= \frac{p_{2}^{2}}{q_{2}^{2}}\left(\bar \eta_2\dot{\bar\eta}_1 -\bar\eta_1\dot {\bar\eta}_2 \right).
\end{eqnarray} 
This completes our analysis of the  connection  between the (anti-)BRST symmetries  and  $W_3$-algebra.
 \section{FCBRST formulation for higher derivative theory}
In this section we investigate the finite coordinate-dependent BRST (FCBRST) formulation for general higher derivative theory. To do so, 
 we first define the infinitesimal BRST symmetry transformation 
 with Grassmannian constant parameter $\delta\rho$ as

 \begin{equation}
 \delta_b q =s q\ \delta\rho,\label{brst}
 \end{equation}
 where $s q$ is the BRST variation of generic variables $q$  in the HD theories. 
The properties of the usual BRST transformation in equation (\ref{brst})  do not depend on whether 
the parameter $\delta\rho$  is (i) finite or infinitesimal, (ii) variable-dependent or not, as long 
as it is anticommuting and global in nature. These observations give us a freedom to 
generalize the BRST transformation by making the parameter $\delta\rho$ finite and coordinate-dependent without
 affecting its properties. We call such generalized BRST transformation in quantum mechanical systems
  as FCBRST transformation.
 In the field theory such generalization is known as FFBRST  transformation \cite{sdj}. 
 Here we adopt a similar technique to generalize the BRST transformation in quantum mechanical theory.
 We start 
by making the  infinitesimal parameter coordinate-dependent with introduction of an arbitrary parameter $\kappa\ 
(0\leq \kappa\leq 1)$.
We allow the generalized coordinates, $q( \kappa)$, to depend on  $\kappa$  in such a way that $q( \kappa =0)=q $ and $q( \kappa 
=1)=q^\prime $, the transformed coordinate.

The usual infinitesimal BRST transformation, thus can be written generically as 
\begin{equation}
{dq( \kappa)}=s  [q  ]\Theta^\prime [q ( \kappa ) ]{d\kappa},
\label{diff}
\end{equation}
where the $\Theta^\prime [q ( \kappa ) ]{d\kappa}$ is the infinitesimal but coordinate-dependent parameter.
The FCBRST transformation with the finite coordinate-dependent parameter then can be 
constructed by integrating such infinitesimal transformation from $\kappa =0$ to $\kappa= 1$, to obtain
\cite{sdj}
\begin{equation}
q^\prime\equiv q (\kappa =1)=q( \kappa=0)+s  (q  )\Theta[q  ],
\label{kdep}
\end{equation}
where 
\begin{equation}
\Theta[q ]=\int_0^1 d\kappa^\prime\Theta^\prime [q( \kappa^\prime)],\label{fin}
\end{equation}
 is the finite coordinate-dependent parameter. 

Such a generalized BRST transformation with finite coordinate-dependent
 parameter is the symmetry  of the effective action in equation (\ref{seff}). However, the 
path integral measure in equation (\ref{zfun}) is not invariant under such transformation as the 
BRST parameter is finite in nature. 
The Jacobian of the path integral measure for such transformations is then evaluated for some 
particular choices of the finite coordinate-dependent parameter, $\Theta[q(x)]$, as
\begin{eqnarray}
{\cal D}q^\prime &=&J(
\kappa) {\cal D}q(\kappa).
\end{eqnarray}
The Jacobian, $J(\kappa )$ can be replaced (within the functional integral) as
\begin{equation}
J(\kappa )\rightarrow \exp[iS_1[q( \kappa) ]],
\end{equation}
 iff the following condition is satisfied \cite{sdj} 
\begin{eqnarray}
 \int {\cal{D}}q   \;  \left [ \frac{1}{J}\frac{dJ}{d\kappa}-i\frac
{dS_1[q (x,\kappa )]}{d\kappa}\right ] 
 \exp{[i(S_{eff}+S_1)]}=0, \label{mcond}
\end{eqnarray}
where $ S_1[q ]$ is local functional of variables such that at $\kappa =0$ it must vanish.

The infinitesimal change in the $J(\kappa)$ is written as \cite{sdj},
\begin{equation}
\frac{1}{J}\frac{dJ}{d\kappa}=-\int d\tau \left [\pm s q ( \kappa )\frac{
\partial\Theta^\prime [q ( \kappa )]}{\partial q ( \kappa )}\right],\label{jac}
\end{equation}
where $\pm$ sign refers to whether $q$ is a bosonic or a fermionic variable.

Thus, the FCBRST transformation with appropriate $\Theta$, changes the
effective action $S_{eff} $ to a  new effective 
action  $S_{eff}+S_1(\kappa=1)$ within the functional integration.  
\section{Connecting different gauges in relativistic particle models}
Here we will exploit the general FCBRST formulation developed in the previous section to connect the path integral of relativistic particle models with different gauge conditions. The FCBRST transformations $(f_b)$ for the relativistic particle model with 
curvature are constructed   as follows:
 \begin{eqnarray}
 f_b q_1^\mu =-c  q_2^\mu \Theta[q],\ \  f_b q_2^\mu =(-\dot c   q_2^\mu -c  \dot q_2^\mu)\Theta[q],\ \   f_b c =0,\ \  f_b\bar c  =B \Theta[q], \ \  f_b B  =0,\label{ffbrst}
 \end{eqnarray}
 where $\Theta[q]$ is an arbitrary finite coordinate-dependent parameter.  
 Now, we show how two different gauges (say $F_1(q) = 0$ and $F_2(q) = 0$) in the relativistic particle model may be connected by such transformations. For this purpose, let us choose the following infinitesimal coordinate dependent parameter (through equation (\ref{fin}))
\begin{equation}
\Theta'[q] =-i\int d\tau\ \bar c (F_1 -F_2 ).
\end{equation}
Let us first calculate the  infinitesimal change in the Jacobian $J(\kappa) $ for above $\Theta'[q]$  using the relation (\ref{jac})  as
\begin{eqnarray}
 \frac{1}{J}\frac{dJ}{d\kappa} &=&i\int d\tau [-B(F_1 -F_2) +s  (F_1 -F_2 ) \bar c ],\nonumber\\
  &=&-i\int d\tau [ B(F_1 -F_2) + \bar c\ s (F_1 -F_2 ) ].\label{jac1}
\end{eqnarray}
To express the Jacobian as  $e^{iS_1}$ \cite{sdj}, we take the ansatz, 
\begin{eqnarray}
S_1 [   \kappa ]=\int d\tau [\zeta_1(\kappa )  B F_1  +\zeta_2 (\kappa ) BF_2    +\zeta_3(\kappa ) \bar c\ s  F_1 +
\zeta_4(\kappa )\bar c\ s  F_2],\label{s1}
\end{eqnarray}
where $\zeta_i(\kappa ) (i =1,...4)$ are constant parameters satisfying the boundary conditions
\begin{equation}
\zeta_i(\kappa =0 ) =0.\label{con}
\end{equation} 
 
To satisfy the crucial condition  (\ref{mcond}), we calculate the
infinitesimal change in $S_1$ with respect to $\kappa$ using the relation (\ref{diff}) as
\begin{eqnarray}
\frac{ dS_1 [ q, \kappa ]}{d\kappa} &=&\int d\tau [\zeta_1'  B F_1  +\zeta_2' BF_2    +\zeta_3' \bar c\ s  F_1 +
\zeta_4'\bar c\ s  F_2\nonumber\\
& +&(\zeta_1 -\zeta_2 )B(s  F_1) \Theta' +(\zeta_2 -\zeta_4 )B  (s  F_2) \Theta'],\label{dis}
\end{eqnarray}
where prime denotes the differentiation with respect to $\kappa$.
Exploiting equations  (\ref{jac1}) and (\ref{dis}), the condition  (\ref{mcond}) simplifies to, 
\begin{eqnarray}
&&\int {\cal{D}}q   \;  \left [ (\zeta_1' +1)  BF_1  +(\zeta_2' -1) BF_2    +(\zeta_3' +1) \bar c\ s  F_1 +
(\zeta_4' -1)\bar c\ s  F_2\right.\nonumber\\
& &+\left.(\zeta_1 -\zeta_3 )B (s  F_1) \Theta' +(\zeta_2 -\zeta_4 )B (s  F_2) \Theta'\right ] 
e^{ i(S_{eff}+S_1) }=0.  
\end{eqnarray}
The comparison of coefficients from the terms of the above equation   
gives the following constraints on the parameters $\zeta_i$
\begin{eqnarray}
&&\zeta_1' +1 = 0,\ \ \zeta_2' -1 =0,\ \ \zeta_3' +1 =0,\ \ \zeta_4' -1 =0,\nonumber\\
&&\zeta_1 -\zeta_3=0,\ \ \zeta_2 -\zeta_4 =0.
\end{eqnarray}
The solutions of the above equations satisfying the boundary conditions (\ref{con}) are
\begin{eqnarray}
 \zeta_1 =-\kappa,\ \ \zeta_2 =\kappa,\ \ \zeta_3 =-\kappa,\ \ \zeta_4 =\kappa.
\end{eqnarray}
With these values of $\zeta_i$  the expression of $S_1[\kappa]$ given in equation (\ref{s1}) becomes
\begin{eqnarray}
S_1 [\kappa ]=\int d\tau [ -\kappa   B F_1  + \kappa  BF_2    - \kappa  \bar c\ s F_1 +
 \kappa  \bar c\ s  F_2],
\end{eqnarray}
which vanishes at $\kappa=0$. 
Now, by adding   $S_1(\kappa =1)$ to the effective action ($S_{eff}$) given in equation (\ref{seff}) we get
\begin{eqnarray}
S_{eff}+S_1(\kappa =1) = S+ \int d\tau \left[ \frac{1}{2} B^2 +BF_2 [f(q) ] +\bar c s F_2 [f(q) ] \right],
\end{eqnarray}
which is nothing but the effective action for relativistic particle models satisfying the
different gauge condition $F_2[f(q)] =0$.
Thus, under FCBRST transformation, the generating functional of HD models
changes from one gauge condition ($F_1 [f(q) ]=0$) to another gauge ($F_2 [f(q) ]=0$) as 
\begin{eqnarray}
\int d\tau e^{iS_{eff}}\stackrel{FCBRST} { ----\longrightarrow}\left( \int d\tau e^{i[S_{eff}+S_1(\kappa =1)]} 
\right).
\end{eqnarray}
We end  this section by noting that the FCBRST transformation with appropriate 
finite coordinate-dependent parameter is able to connect two different (arbitrary) gauges of the relativistic particle model.   
  \section{Conclusions}
The  relativistic particle models have always been an interesting area of research as it led to the Polyakov action of string theory \cite{polya}. When a curvature term is added to the action of the relativistic particle model it becomes a higher derivative (HD) theory. Due to HD nature, it shows an inconsistency in counting the independent gauge degrees of freedom. The apparent mismatch is due to the interrelation between the variables with higher derivatives. Whereas, if we consider the mass term to be zero (with proper condition on the particle velocities as in \cite{plyuschay2}) the mismatch vanishes and the number of gauge degrees of freedom and number of independent primary first class constraints are same \cite{BMP}, as happens for all standard theories \cite{BRR, HRT}. So, it would be interesting to study the BRST symmetries of both these models. But here we are faced an obstacle. For HD theories there is no well defined prescription for analysing BRST symmetry. In the present case this is avoided by working in the first order formalism developed in \cite{BMP}.  

   In this paper, we have analysed the different constraint structures of
 the models  of relativistic particle  with curvature
   and  of massless relativistic particle with rigidity. The relativistic particle model with curvature is shown to have the diffeomorphism symmetry whereas the gauge symmetries of the model of  relativistic massless particle  with rigidity
   contain both diffeomorphism and $W$-morphisms.   
  The nilpotent BRST and anti-BRST symmetries for these model have also been
  investigated. A remarkable feature for such symmetries is the manifestation of
  $W_3$-algebra. The BRST symmetries for all  variables (excluding for anti-ghost variable)
   corresponding to diffeomorphism and $W$-morphism satisfy the $W_3$-algebra.
  Likewise, apart from the ghost variable, the anti-BRST symmetry transformations for all other variables  
  also satisfy the same $W_3$-algebra. Thus the full $W_3$-algebra for all  variables is obtained by taking into account both BRST and anti-BRST transformations.
  
  The finite coordinate-dependent BRST (FCBRST) symmetry,
  which is quantum mechanical analog of finite field-dependent BRST (FFBRST), has also been analysed in full generality for higher derivative particle models.
 It has been shown that although such a transformation is a symmetry of the effective action, it  breaks the
 invariance of the generating functional of the path integral. The Jacobian of path integral measure
 changes non-trivially for FCBRST symmetry transformation. We have shown that 
 FCBRST transformation with a suitable coordinate-dependent parameter changes the
effective action from one gauge to another within a functional integral. Thus,
FCBRST formulation is very useful to
connect two different Greens functions for models of relativistic particles. The results were explicitly presented for the massive case. For the massless version, all results go over trivially in the appropriate limit. Finally, we feel that although our analysis was done for relativistic particle models, it is general enough to include other higher derivative models.
 
 \section*{Acknowledgement}
 One of the authors  (BP)    gratefully acknowledges the Council of Scientific and Industrial Research (CSIR), Government of India, for financial assistance.


\begin{thebibliography}{999}
 \bibitem{podolsky1} B.~Podolsky, Phys.\ Rev.\  {\bf 62} (1942) 68.
\bibitem{podolsky2} B.~Podolsky,
 C.~Kikuchi,
Phys.\ Rev.\  {\bf 65}  (1944) 228.
  \bibitem{Iliopoulos} J. Iliopoulos, B. Zumino, Nucl.\ Phys.\  B \textbf{76} (1974) 310.
  \bibitem{Gama} F. S. Gama, M. Gomes, J. R. Nascimento, A. Yu. Petrov,  A. J. da Silva. Phys. Rev. D \textbf{84}  (2011) 045001.
   \bibitem{clz} C.~S.~Chu, J.~Lukierski, W.~J.~Zakrzewski, Nucl.\ Phys.\  B {\bf 632}  (2002) 219.
    \bibitem{plyuschay6} P. D. Alvarez, J. Gomis, K. Kamimura, M. S. Plyushchay, Phys. Lett. B \textbf{659} (2008) 906 [arXiv:0711.2644].
   \bibitem{neupane} I. P. Neupane, JHEP  {\bf 09} (2000) 040.  
\bibitem{nojiri4} S. Nojiri, S. D. Odintsov, S. Ogushi, Phys. Rev. D \textbf{65}  (2001) 023521.
\bibitem{reyes} C. ~M. ~Reyes, Phys. Rev D \textbf{80}, 105008 (2009).
\bibitem{MP}  P. ~Mukherjee, B. ~Paul, Phys. Rev. D \textbf{85} (2012) 045028.
\bibitem{plyuschay3} M. S. Plyushchay,  Nucl. Phys. B \textbf{362 }(1991)
54.
\bibitem{plyuschay4} Peter A. Horv\' athy, M. S. Plyushchay, JHEP \textbf{0206} (2002) 033 [hep-th/0201228]
\bibitem{plyuschay5} M. S. Plyushchay, Electron. J. Theor. Phys. \textbf{3N10}
(2006) 17 [math-ph/0604022].
\bibitem{plyuschay7} M. S. Plyushchay, Phys. Lett. B \textbf{262} (1991) 71.
\bibitem{cordero} R. Cordero, A. Molgado, E. Rojas, Class. Quantum Grav. \textbf{28} (2011) 065010.
\bibitem{paul} B. Paul, Phys. Rev. D \textbf{ 87} (2013) 045003.
\bibitem{stelle} K.~S. ~Stelle,  Phys. Rev. D {\bfseries 16} (1977) 953.
\bibitem{accioly} A. J. Accioly, Revlsta Brasllelra de Flslca \textbf{18 } (1988) 593.
\bibitem{soti} T. P. Sotiriou, V. Faraoni, Rev. Mod. Phys. \textbf{82} (2010) 451 [arXiv:0805.1726].
 \bibitem{gullu} I. Gullu, T. C. Sisman, B. Tekin, Phys. Rev. D \textbf{81} (2010) 104017.
 \bibitem{ohta} N. Ohta, Class. Quantum Grav. \textbf{29} (2012) 015002.
 \bibitem{polya} A. M. Polyakov, Nucl. Phys. B \textbf{268} (1986) 406.
\bibitem{elie} D. A. Eliezer, R. P. Woodard, Nucl.\ Phys.\ B \textbf{325} (1989) 389.
\bibitem{pisarski} R. D. Pisarski, Phys. Rev. D {\bf{34}} (1986)  670.
\bibitem{nesterenko} V. V. Nesterenko, J. Phys. A {\bf{ 22}} (1989) 1673.
\bibitem{plyuschay1}M. S. Plyushchay,  Mod. Phys. Lett. A \textbf{3} (1988) 1299;
 \bibitem{plyuschayA}M. S. Plyushchay, Int. J. Mod.Phys. A \textbf{4} (1989) 3851.
\bibitem{plyuschay2} M. S. Plyushchay, Mod. Phys. Lett. A \textbf{4} (1989) 837, Phys. Lett. B {\bf 243} (1990) 383.

\bibitem{ramos} E. Ramos, J. Roca, Nucl. Phys. B {\bf{ 436}} (1995)  529.
\bibitem{ramos2} E. Ramos, J. Roca, Nucl. Phys.  B \textbf{452 } (1995)  705. 
\bibitem{BMP} R. ~Banerjee, P. ~Mukherjee, B. ~Paul, JHEP {\bf 1108} (2011) 085.
\bibitem{novikov} S. Novikov, S. V. Manakov, L. P. Pitaevskii, V. E. Zakharov, \textit{Theory of solitons: The inverse Scattering Method}, Contemporary Soviet Mathematics, Consultants Bureau [Plenum], New York, 1984.
 \bibitem{brst} C. Becchi, A. Rouet, R. Stora, Annals Phys. {\bf{98}} (1974) 287.
\bibitem {tyu}I. V. Tyutin, LEBEDEV-{\bf 75-39} (1975).
\bibitem{ht} M. Henneaux, C. Teitelboim, {\it{ Quantization of gauge
systems}}, Princeton, USA: Univ. Press (1992).
\bibitem{wei} S. Weinberg, {\it{ The quantum theory of fields, Vol-II: Modern
applications}}, Cambridge, UK Univ. Press (1996).
\bibitem{sdj} S. D. Joglekar,  B. P. Mandal,  Phys. Rev. D {\bf 51}  (1995)  1919. 

\bibitem{sb1} S. D. Joglekar, B. P. Mandal, Int. J. Mod. Phys. A \textbf{17},  (2002) 1279. 
\bibitem{jm} S. D. Joglekar, A. Misra, J. Math. Phys \textbf{41}, (2000) 1755.
 \bibitem{ssb}  B. P. Mandal,  S. K. Rai,  S.  Upadhyay, Eur. Phys. Lett. { \bf 92}  (2010) {21001}.
 \bibitem{susk}   S. Upadhyay,   S. K. Rai,  B. P. Mandal,  J. Math. Phys.  {\bf 52}  (2011) {022301}.
\bibitem{sb} S. Upadhyay, B. P. Mandal,  Mod. Phys. Lett.   A {\bf 40}  (2010)  { 3347}; Eur. Phys. Lett.  {\bf 93}  (2011) 
{31001}; Eur. Phys. J.   C {\bf 72},   2059 (2012);   Eur. Phys. J.  C {\bf 72}  (2012) 2065; Annals of Physics { \bf 327}  (2012) 28850. 

\bibitem{rb}  R. Banerjee, B. P. Mandal,  Phys. Lett.     B {\bf 27}  (2000) {488}.
\bibitem{smm} S. Upadhyay, M. K. Dwivedi, B. P. Mandal, Int. J. Mod. Phys. A {\bf 28}  (2013) 1350033.
\bibitem{fs} M. Faizal, B. P. Mandal, S. Upadhyay, Phys. Lett. B {\bf 721}  (2013) 159.

\bibitem{ostro} M. Ostrogradsky, \textit{Mem. Ac. St. Petersbourg} {\bf V14} (1850) 385.
\bibitem{BRR} R.~Banerjee, H.~J.~Rothe,  K.~D.~Rothe, Phys. Lett. B {\bf{463}} (1999) 248, [hep-th/9906072];
 Phys. Lett. B {\bf{479}} (2000) 429, [hep-th/9907217];  
  J. Phys. A \textbf{33} (2000) 2059;  [hep-th/9909039].  
\bibitem{HRT} A.~Hanson, T.~Regge, C.~Tietelboim, {\it Constrained Hamiltonian System}, (Accademia Nazionale Dei Lincei, Roma, 1976).

\end{thebibliography}
\end{document}